\documentclass[a4paper]{article}
\usepackage{amssymb}
\usepackage{fleqn}
\usepackage{epsfig,a4}

\let \widehat=\hat

\let\kappa=\varkappa

\def\text#1{\mbox{#1}}

\def\restylefloat#1{}
\newcommand{\ds}{\displaystyle}
\newcommand{\abc}[1]{\mbox{#1)}\quad}
\newcommand{\ø}{{\!}}
\newcommand{\bm}[1]{\mbox{\boldmath $#1$}}
\newcommand{\grad}{\mathop{\rm grad}\nolimits}

\newcommand{\deriv}[2]{\mbox{$\displaystyle
\frac{\mathrm{d}\,\! #1}{\mathrm{d}\,\! #2}$}}
\newcommand{\dd}{\mathrm{d}\,\!}

\begin{document}  \thispagestyle{plain}

\title{Dark Matter and Rotation Curves of Stars in Galaxies}
\author{E. Schmutzer\thanks{eschmu@.aol.com}, Jena, Germany \\
Friedrich Schiller University}
\date{Received: 2001}
\maketitle

\begin{abstract}
The dark matter accretion theory (around a central body) of the author on
the basis of his 5-dimensional Projective Unified Field Theory (PUFT) is
applied to the orbital motion of stars around the center of the Galaxy. The
departure of the motion from Newtonian mechanics leads to approximately flat
rotation curves being in rough accordance with the empirical facts. The
spirality of the motion is investigated.
\end{abstract}

\section{Retrospect}

In a preceding paper (Schmutzer 2001) we developed a theory of dark matter
accretion around a central spherically symmetric body on the basis of PUFT
(Schmutzer 1995, 2000a, 2000b, 2000c) which we applied to the motion of a
test body around a central body with the result of an additional radial
acceleration effect towards the center, compared with the motion on the
basis of Newtonian mechanics. Particularly this effect was used to explain
the empirically measured deviation of the motion of the satellites Mariner
10/11 (Pioneer effect) from the Newtonian motion.

In this paper we apply our dark matter accretion theory to the orbital
motion of the stars around the center of the Galaxy with the aim to find an
explanation of the approximetely flat rotation curves of the orbiting stars.

In this context we mention interesting papers of Dehnen et al. (Dehnen, Rose
and Amer 1995) who elaborated a theory of the flat rotation curves on a
different theoretical basis applying the Bose-Einstein statistics (preferred
by physical arguments) and the Fermi-Dirac statistics (dropped by physical
arguments). By these authors quantum-statistical calculations were performed
in great detail.

For a rough understanding of our approach we restricted our investigations
to the Boltzmann-Maxwell statistics describing a gas of one sort of dark
matter particles (dm-particles) without degeneracy. We think this step is
legitimate to simplify the sitation, with the advantage to be able to
perform the calculations fully analytically.

According to this dark matter accretion theory, applied to a spherical
homogeneous central body following relationship between the gravitational
potential $\chi (r)$ and the dm-particle number density $n(r)$ reads:

\begin{equation}
n=-\frac{\bar{n}m\chi }{\text{k}T}  \label{eins}
\end{equation}
($\bar{n}$ dm-particle number density at ``infinity'' (great
distance), $m$ mass of a dm-particle, $T$ kinetic temperature of
the gas, k Boltzmann constant).

The potential obeys the differential equation ($\gamma _{N}$ Newtonian
gravitational constant)

\begin{equation}
\Delta \chi +\kappa ^{2}\chi =4\pi \gamma _{N}\widehat{\mu }\, ,
\label{zwei}
\end{equation}
where with respect to the mass density the quantity $\widehat{\mu }$ means $%
\mu _{0}$ (mass density in the interior) and $\mu _{G}$ (mass density in the
exterior). Further $\kappa ^{2}$ is defined by ($\kappa $ dark matter
parameter)

\bigskip
\begin{equation}
\kappa ^{2}=\frac{4\pi \gamma _{N}m^{2}\bar{n}}{\text{k}T}\, .
\label{drei}
\end{equation}

One should realize the positive sign in front of the second term on the left
hand side of equation (\ref{zwei}). Let us in this context mention that we
arrived at this approach to our theory of dark matter accretion by our
experience in the Debye-Milner theory of strong electrolytes, where despite
attractive forces between negative and positive ions at this place a minus
sign appears which leads to a Yukawa-type potential, whereas in our theory
periodic functions occur.

For presenting the gravitational potentials etc. in the interior and in the
exterior of a homogeneous sphere considered following two basic functions
are useful:
\begin{equation}\label{vier}
\begin{array}{lll}\bigskip
\abc{a}U(z) &=&z\sin z+\cos z\, ,
   \\
\abc{b}V(z) &=&\sin z-z\cos z
\end{array}
\end{equation}
which determine the potentials as it will be shown in the following.

For the gravitational potential and the corresponding derivatives
we received for the interior (index $i$) and the exterior (index
$e$) of the sphere the results ($r_{0}$ radius of the sphere)
\begin{equation}\label{fuenf}
\begin{array}{lll}\ds
\bigskip
 \abc{a}\chi _{i} &=&\ds B_{0}\{1-\frac{1}{\kappa r}U(\kappa r_{0})\sin
(\kappa r)\}\, ,
\\ \ds
\abc{b}\deriv{\chi _{i}}{r} &=&\ds\frac{B_{0}}{\kappa r^{2}}\,
 U(\kappa r_{0})V(\kappa r)\, ;
\end{array}
\end{equation}
\begin{equation} \label{sechs}
\begin{array}{lll}  \ds \bigskip
\abc{a}\chi _{e} &=&\ds-\frac{B_{0}}{\kappa r}\,V(\kappa
r_{0})\cos (\kappa r)\, ,  \\  \ds \abc{b}\deriv{\chi _{e}}{r}
&=&\ds\frac{B_{0}}{\kappa r^{2}}\, V(\kappa r_{0})U(\kappa r)\, .
\end{array}
\end{equation}
The constant of integration $B_{0}$ occurring in these formulas is given
by\bigskip\ \ \qquad\ \qquad \qquad \qquad
\begin{equation}
B_{0}=\frac{\mu _{0}\text{k}T}{m^{2}\bar{n}}=\frac{3\gamma _{N}M_{c}}{%
\kappa ^{2}r_{0}^{3}}\, ,  \label{sieben}
\end{equation}
where
\begin{equation}
M_{c}=\frac{4\pi \mu _{0}r_{0}^{3}}{3}  \label{acht}
\end{equation}
($M_{c}$ mass of the central body) holds.

Let us for application in the following mention the equation of motion of a
test body

\begin{equation}
\deriv{\bm{v}}{t}+\grad\chi +\bm{v}\deriv{\ln \sigma _{c}}{t}=0 \,
, \label{neun}
\end{equation}
where $\bm{v}$ is the velocity of the test body and $\sigma _{c}$
is the cosmological scalaric world function.

\section{Motion of a test body in polar coordinates}

Now we use the polar coordinatets \{$R,$ $\Phi $ \}. Then after integration
the equation (\ref{neun}) reads (dot means derivative with respect to the
time $t$)
\begin{equation} \label{zehn}
\begin{array}{l}  \ds \bigskip
\abc{a}  \ddot{R}-R\,\dot{\Phi }^{2}+ \deriv{\chi
}{R}+\dot{R}\,\Sigma =0,\\ \ds \bigskip \abc{b}
R^{2}\dot{\Phi}\sigma _{c}=\left| \bm{r} \times \bm{v}\right|
\sigma _{c}\equiv F_{0}\quad\text{with}
\\ \ds
\abc{c}\Sigma =\deriv{ln\sigma _{c}}{t} \, .
\end{array}
\end{equation}
As it is well known, the azimuthal velocity is defined by
\begin{equation}
v_{\Phi }=R\stackrel{\cdot }{\Phi }=\frac{F_{0}}{R\sigma _{c}}\, .
\label{elf}
\end{equation}
With the aid of this expression formula (\ref{zehn}a) leads to

\begin{equation}
\stackrel{\cdot \cdot }{R}-\frac{v_{\Phi }^{2}}{R}+\frac{\dd \chi }{%
\dd R}+\stackrel{\cdot }{R}\Sigma =0\, .  \label{zwoelf}
\end{equation}
Hence the equation

\begin{equation}
v_{\Phi }=\sqrt{R \,\ddot{R}+R\,\dot{R}\,\Sigma +
R\deriv{\chi }{R}}  \label{zehndrei}
\end{equation}
follows. Now we introduce the abbreviation

\begin{equation}
A=\ddot{R}+\dot{R}\,\Sigma\, .
\label{zehnvier}
\end{equation}
In the usual theory of the rotation curves on the basis of Newtonian
mechanics approximately the motion of the body on a circle ($R=const$) is
considered. We think this approximation is not good enough. Therefore,
induced by the empirical situation for small values of the radial coordinate
$R$ we try the basical ansatz

\begin{equation}
A=\alpha R=10^{-30}aR\, ,  \label{zehnfuenf}
\end{equation}
where $\alpha $ or $a$ (the factor $10^{-30}$ is an adaption factor with
respect to the large numerical astronomical values) is a quantity to be
specified by further concrete physical assumptions . Substituting this
expression into (\ref{zehndrei}) gives the formula for the so-called
rotation curve $V=v_{\Phi }(R)$:

\begin{equation}
V=\sqrt{R\left(A+\deriv{\chi }{R}\right)}\, .  \label{zehnsechs}
\end{equation}
Finishing this section we would like to emphasize that in our theory the
situation is basically different to the usual theory of the rotation curves
based on the assumption $R=const$, since here because of the periodicity
occurring in the potential the expression $\ds\deriv{\chi }{R}$
is indefinite, i.e. that for $A=0$ the radicand exhibits negativ-valued
regions.

\section{Rotation curve for the interior and exterior of the central body in
general (azimuthal motion)}

\textbf{Interior} (index $i$):

With the help of (\ref{fuenf}b) and (\ref{sieben}) equation (\ref{zehnsechs})
takes the shape (approximation by power series expansion with respect to ($
\kappa R)^{2}\ll 1$)

\begin{equation}
V_{i}=R\sqrt{\ds\alpha _{i}+\frac{3\gamma _{N}M_{c}}{(\kappa r_{0})^{3}R^{3}}\,
U(\kappa r_{0})V(\kappa R)}\,.  \label{zehnsieben}
\end{equation}
Hence for the angular velocity (differential rotation) follows

\begin{equation}
\omega _{i}=\frac{V_{i}}{R}=\sqrt{\alpha _{i}+\frac{3\gamma _{N}M_{c}}{%
(\kappa r_{0})^{3}R^{3}}\,U(\kappa r_{0})V(\kappa R)}\,.  \label{zehnacht}
\end{equation}

\textbf{Exterior} (index $e$):

In this case with the help of (\ref{sechs}b) and (\ref{sieben}) equation (%
\ref{zehnsechs}) takes the shape

\begin{equation}
V_{e}=R\sqrt{\alpha _{e}+\frac{3\gamma _{N}M_{c}}{(\kappa
r_{0})^{3}R^{3}}\, V(\kappa r_{0})U(\kappa R)}\, .
\label{zehnneun}
\end{equation}

Analogously follows here the angular velocity

\begin{equation}
\omega _{e}=\frac{V_{e}}{R}=\sqrt{\alpha _{e}+\frac{3\gamma _{N}M_{c}}{%
(\kappa r_{0})^{3}R^{3}}\,V(\kappa r_{0})U(\kappa R)}\, .  \label{zwanzig}
\end{equation}

\section{Physical processes in the interior of the central body (radial
motion)}

Our next task is to get some information about the quantity $A$ in the
relation (\ref{zehnsechs}) by investigating physical main processes in the
interior of the central body. We propose to approach this problem
tentatively by considering two opposite processes:
\begin{enumerate}
\item Expansion process of the central body by scalaric cosmological bremsheat
production according to our hypothesis (Schmutzer 2000c).
\item Scalaric-adiabatic cosmological contraction process according to our
approximate treatment of the orbital motion of a test body considered
(Schmutzer 2000a).
\end{enumerate}

\subsection{Expansion process (radial motion)}

In the paper quoted above we started from the radial expansion formula of a
sphere ($r$ radius of the sphere)

\begin{equation}
\dd r_{Q}=\frac{1}{3}\alpha _{c}r\dd T  \label{zwanzigeins}
\end{equation}
($T$ temperature of the body, $\alpha _{c}$ cubic heat expansion
coefficient, index $Q$ refers to heat). Further the relationship

\begin{equation}
\dd Q=W\dd T  \label{zwanzigzwei}
\end{equation}
holds ($Q$ produced heat), where for the heat capacity

\begin{equation}
W=c_{Q}M  \label{zwanzigdrei}
\end{equation}
($M$ mass of the central body considered here, $c_{Q}$ specific heat of the
body) is valid.

Let us now tentatively identify the heat $Q$ with the scalaric cosmological
bremsheat $Q_{B}$ produced by the bremsheat effect predicted by us:

\begin{equation}
\dd Q = f_{S}\dd Q_{B}  \label{zwanzigvier}
\end{equation}
($f_{S}$ free scalaric consumption factor). Two years ago we derived for the
differential of the bremsheat production the expression

\begin{equation}
\dd Q_{B}=\dd Q_{B}^{(transl)}+\dd Q_{B}^{(rot)}=
\left[ M \,\dot{\bm{r}}_{C}^{2}+\sum\limits_{a,b}
I_{ab}\omega _{a}\omega _{b}\right] \dd \ln \sigma _{c}
\label{zwanzigfuenf}
\end{equation}
($M$ mass of a body with translational and rotational motion,
$\bm{r}_{C} $ radius vector to the center of mass of the body
considered, $I_{ab}$
moment of inertia tensor, $\omega _{a}$ angular velocity, indices $%
a,b=1,2,3).$ For a massive sphere rotating around a fixed axis results

\begin{equation}
\dd Q_{B}=\left[ M\,\dot{\bm{r}}_{C}^{2}+ \frac{2M}{5}(r_{0}\omega
)^{2}\right] \dd \ln \sigma _{c} \label{zwanzigsechs}
\end{equation}
Using the abbreviation

\begin{equation}
J=M\,\dot{\bm{r}}_{C}^{2}+\frac{2M}{5}(r_{0}\omega
)^{2} \label{zwanzigsieben}
\end{equation}
we can simplify (\ref{zwanzigsechs}):

\begin{equation}
\dd Q_{B}=J\dd\ln \sigma _{c}\, .  \label{zwanzigacht}
\end{equation}
Then by means of (\ref{zwanzigvier}) from (\ref{zwanzigzwei}) results
\begin{equation}
\dd T=\frac{Jf_{S}}{W}\dd \ln \sigma _{c}\,.  \label{zwanzigneun}
\end{equation}
Now it is convenient to introduce the scalaric-thermal velocity coefficient

\begin{equation}
\alpha _{Q}=\frac{Jf_{S}\alpha _{c}\Sigma }{3W}\, .
\label{dreissig}
\end{equation}
Then with the help of (\ref{zwanzigneun}) we receive from (\ref{zwanzigeins})
the bremsheat induced radial velocity

\begin{equation}
v_{rQ}\equiv \frac{\dd r_{Q}}{\dd t}=\alpha _{Q}r\,.
\label{dreissigeins}
\end{equation}
Here one should remember that the coefficent $\alpha _{Q}$ as well as other
cosmological quantities exhibit a slight scalaric-adiabatic cosmological
time dependence.

\subsection{Scalaric-adiabatic contraction process (radial motion)}

As quoted above, some years ago we treated the circle-like orbital motion of
a test body, arriving on the basis of our scalaric-adiabatic approximation
at a contraction process with following result for the orbital radius of the
moving body:

\begin{equation}
r_{S}(t)=\frac{F_{0}^{2}}{\gamma _{N}\mathcal{M}_{c}\sigma
_{c}(\sigma _{c}^{2}-1)}\,.  \label{dreissigzwei}
\end{equation}
Hence we get by differentiation the expression

\begin{equation}
\dot{r}_{S}(t)=-\frac{F_{0}^{2}(3\sigma _{c}^{2}-1)\Sigma }{
\gamma _{N}\mathcal{M}_{c}\sigma _{c}(\sigma _{c}^{2}-1)^{2}}\, .
\label{dreissigdrei}
\end{equation}
We would like to remind the reader that according to PUFT the true constant
mass of a body is the (invariant) mass $\mathcal{M}$ (calligraphic) which is
related to the (inertal) mass $M$ and to the (gravitative-scalaric) mass
$M^{(S)}$ as follows (here $\sigma $ is the general scalaric field function):
\begin{equation} \label{dreissigvier}
\begin{array}{llll}  \bigskip\ds
\abc{a} M &=&\mathcal{M}\sigma \, ,   \\ \bigskip\ds \abc{b}
M^{(S)} &=& \mathcal{M}\sigma ^{(S)}\quad &\text{with}   \\
\bigskip\ds
 \abc{c}\sigma ^{(S)} &=&\ds\frac{\sigma ^{2}-1}{\sigma}
\quad & \text{(scalaric force correction factor).}
\end{array}
\end{equation}
As pointed out in our publications, since about 1000 years after the big
start the inequality $\sigma _{c}^{2}\gg 1$ is well justified, i.e. the
equations (\ref{dreissigzwei}) and (\ref{dreissigdrei}) read much simpler:

\begin{equation}
r_{S}=\frac{F_{0}^{2}}{\gamma _{N}\mathcal{M}_{c}\sigma _{c}^{3}}
\label{dreissigfuenf}
\end{equation}
and

\begin{equation}
\dot{r}_{S}=-\frac{3F_{0}^{2}}{\gamma _{N}\mathcal{M}_{c}
\sigma_{c}{}^{4}}\dot{\sigma}_{c}\, .
\label{dreissigsechs}
\end{equation}
By differentiation we find the expression

\begin{equation}
\ddot{r}_{S}=-\frac{3F_{0}^{2}}{\gamma _{N}
\mathcal{M}_{c}\sigma _{c}\ø^{4}}
\left(\ddot{\sigma}_{c}-
\frac{4}{\sigma }\dot{\sigma}_{c}^{2}\right)\, .
\label{dreissigsieben}
\end{equation}
Substituting (\ref{dreissigfuenf}) till (\ref{dreissigsieben}) into the
relation (\ref{zehnvier}), being applied to this case, yields for the
rearranged formula (\ref{zehnfuenf}) the result

\begin{equation}
\alpha _{S}=\frac{A_{S}}{r_{S}}=
-\frac{3}{\sigma _{c}}\left( \ddot{\sigma}_{c}-
\frac{3}{\sigma _{c}}\dot{\sigma}_{c}^{2}\right)\, .
\label{dreissigacht}
\end{equation}
For eliminating the second order derivative of the scalaric field function
$\sigma _{c}$ we have to go back to the corresponding field equation
contained in our publications (Schmutzer 2000b), namely ($L$ rescaled radius
$K$ of the cosmos, $\eta $ rescaled time $t$, $\vartheta$ energy density of
the substrate in the cosmos, $A_{0}=10^{27}$cm rescaling factor,
$\kappa _{0}$ Einsteinian gravitational constant)

\begin{equation}  \label{dreissigneun}
\begin{array}{l}  \bigskip\ds
\abc{a}
\deriv{^{2}\sigma _{c}}{\eta ^{2}} =3\hat{\kappa}\vartheta
-\frac{3}{L}\, \deriv{L}{\eta }\,
\deriv{\sigma_{c}}{\eta }\quad \text{with}   \\    \ds
\abc{b}
\hat{\kappa} =\frac{\kappa _{0}A_{0}^{2}}{6}
=3.4617\cdot 10^{5}\mathrm{g^{-1}cm\, s}^{2}\, .
\end{array}
\end{equation}
In this context one should remember the rescaling relation

\begin{equation}
\deriv{t}{\eta }=\frac{1}{3}\cdot 10^{17}\mathrm{s}\,.
\label{vierzig}
\end{equation}
After this recalculation the formula (\ref{dreissigacht}) reads
\begin{equation}
\alpha _{S}=
\frac{9}{\sigma _{c}}\,
\left( \deriv{\eta }{t}\right)^{2}
\left[\frac{1}{\sigma _{c}}
\left( \deriv{\sigma _{c}}{\eta }\right)^{2}+
\frac{1}{L}\deriv{L}{\eta }\deriv{\sigma _{c}}{\eta }-
\hat{\kappa}\vartheta \right]\, .  \label{vierzigeins}
\end{equation}
Finishing this subsection we eliminate $F_{0}^{2}$ in (\ref{dreissigsechs})
by means of (\ref{dreissigfuenf}) and obtain

\begin{equation}
\dd r_{S}=-3r\dd \ln \sigma _{c}\,.  \label{vierzigzwei}
\end{equation}

\subsection{Superposition of both radial effects}

In both preceding subsections we treated in detail the expansion effect (\ref
{zwanzigeins}) and the contraction effect (\ref{vierzigzwei}). By adding
both effects we arrive at the total effect for the interior

\begin{equation}
\dd r_{i}=\dd r_{Q}+\dd r_{S}\,.  \label{vierzigdrei}
\end{equation}
Introducing the abbreviation

\begin{equation}
\tilde{\alpha}=\alpha _{Q}-3\Sigma  \label{vierzigvier}
\end{equation}
leads us to the relationship between the radius $r$ and the total radial
velocity

\begin{equation}
\dot{r}_{i}=\tilde{\alpha }r_{i}\,. \label{vierzigfuenf}
\end{equation}
For further calculations it is convenient to use the scalaric bremsheat
parameter

\begin{equation}
\beta _{Q}=\frac{Jf_{S}\alpha _{c}}{3W}\, .  \label{vierzigsechs}
\end{equation}
Then (\ref{dreissig}) and (\ref{vierzigvier}) take the form
\begin{equation} \label{vierzigsieben}
\begin{array}{lll}\bigskip\ds
\abc{a}\alpha _{Q} &=&\beta _{Q}\Sigma\quad \text{and}
\\    \ds
\abc{b}\tilde{\alpha} &=&(\beta _{Q}-3)\Sigma\, .
\end{array}
\end{equation}
For the following we specialize our considerations, having led to relation
(\ref{zwanzigsechs}), to the central body $(M\rightarrow M_{c})$ and further
particularly to the case of a central body with a resting center of mass
($\stackrel{\cdot }{\bm{r}_{C}}=0).$ Then formula
(\ref{zwanzigsieben}) reduces to

\begin{equation}
J=\frac{2M_{c}}{5}(r_{0}\omega )^{2}\,.  \label{vierzigacht}
\end{equation}
Neglecting the scalaric-adiabatic time dependence of the material
constitution quantity $\beta _{Q}$, then by means of (\ref{vierzigsieben}b)
we arrive by differentiation of (\ref{vierzigfuenf}) at the relation

\begin{equation}
\ddot{r}_{i}=(\beta _{Q}-3)
\left[\dot{\Sigma}+(\beta _{Q}-3)\Sigma ^{2}\right]r_{i}\, .
\label{vierzigneun}
\end{equation}

Inserting the expressions (\ref{vierzigfuenf}) and (\ref{vierzigneun}) into
the formula (\ref{zehnvier}), applied to the interior, leads to
($R\rightarrow r_{i}$)

\begin{equation}
A_{i}=\alpha _{i}r_{i}  \label{fuenfzig}
\end{equation}
with

\begin{equation}
\alpha _{i}=(\beta _{Q}-3)\left[ \dot{\Sigma}+
(\beta_{Q}-2)\,\Sigma ^{2}\right]\, .  \label{fuenfzigeins}
\end{equation}

\section{Physical processes in the exterior of the central body (radial
motion)}

The difference between the exterior and the interior consists in the fact
that in the exterior no bremsheat effect takes place. This means that a
moving body firstly, when leaving by jet the central body, starts with the
constant initial velocity

\begin{equation}
\dot{r}_{jet}=\dot{r}_{Q}(r=r_{0})=\alpha_{Q}\,r_{0}\, ,
\label{fuenfzigzwei}
\end{equation}
and secondly underlies the scalaric-adiabatic cosmological contraction (\ref
{vierzigzwei}). Superposition of both effects gives
\begin{equation}
v_{e}\equiv \dot{r}_{e}=\alpha _{Q}r_{0}-3r\Sigma \,.
\label{fuenfzigdrei}
\end{equation}
In this case according to (\ref{zehnvier}) follows

\begin{equation}
A_{e}=\ddot{r}_{e}+\dot{r}_{e}\Sigma =\alpha_{e}r\, ,
\label{fuenfzigvier}
\end{equation}
where

\begin{equation}
\alpha _{e}=\Lambda -3\Sigma ^{2}+\beta _{Q}\Sigma ^{2}\frac{r_{0}}{r}
\label{fuenfzigfuenf}
\end{equation}
with
\begin{equation}
\Lambda =-3\left[ \dot{\Sigma}-3\Sigma^{2}\right]
\label{fuenfzigsechs}
\end{equation}
holds.

\section{Orbital motion of a test body in the exterior of the central body}

\subsection{Rearrangement of the velocity equations in usual astronomical
units}

In presenting the plots of the rotation curve and the spiral motion, in
astrophysics the azimuthal velocity of the moving body (in astrophysics the
notion ``rotation velocity'' is used) is indicated in
$\ds \frac{\mathrm{km}}{\mathrm{s}}$, whereas its distance from the central body is given in kpc. In
adapting to this practice we have to rearrange the equations (\ref{zehnneun})
and (\ref{fuenfzigdrei}), using the relationship  $1\mathrm{kpc = 3.086\cdot 10^{16}\,km}$

The result of this recalculation is the following:

\begin{equation}
V_{e}=30.86\cdot \mathrm{ km}\cdot R\cdot \mathrm{(kpc)}^{-1}
\sqrt{a_{e}+\frac{3\gamma _{N}M_{c}\cdot 10^{30}}{(\kappa r_{0})^{3}
R^{3}}\,V(\kappa r_{0})U(\kappa R)}  \label{fuenfzigsieben}
\end{equation}
and ($r\rightarrow R)$

\begin{equation}
v_{e}=30.86\cdot 10^{15}\cdot \mathrm{km}\cdot \left[ \alpha
_{Q}r_{0}-3R\Sigma \right]\cdot \mathrm{(kpc)}^{-1}\, .
\label{fuenfzigacht}
\end{equation}

Both equation are the basis for our numerical investigations, particularly
for plotting the rotation curves and the spirals, that will be done later.

\subsection{Spirality of the motion of the test body in the exterior}

The spiral form of the orbiting test body results by superposition of the
radial motion, described by the radial velocity (\ref{fuenfzigacht}), and
the azimuthal velocity (\ref{fuenfzigsieben}). Let us introduce the notion
``spirality'' by the definition

\begin{equation}
Sp=\frac{V_{e}}{v_{e}}=\tan \chi \, ,  \label{fuenfzigneun}
\end{equation}
where $\chi $ will be called spirality angle. Then the final formula for the
spirality reads

\begin{equation}
Sp=\tan \chi =
\frac{\sqrt{a_{e}+\ds\frac{3\gamma _{N}M_{c}\cdot 10^{30}}
{(\kappa r_{0})^{3}R^{3}}\,
V(\kappa r_{0})U(\kappa R)}}{10^{15}\left(\alpha _{Q}\ds
\frac{r_{0}}{R}-3\Sigma \right)}\, .
\label{sechzig}
\end{equation}
From this expression we learn that the reversion point of the radial motion
is determined by the quantity

\begin{equation}
R_{rev}=\frac{\alpha _{Q}r_{0}}{3\Sigma }\,.  \label{sechzigeins}
\end{equation}
Remembering the relation

\begin{equation}
\abc{a}v_{e}=\dot{R}\qquad\text{and}\qquad
\abc{b}V_{e}=R\dot{\Phi}
\label{sechzigzwei}
\end{equation}
in polar coordinates, we arrive at the following relationship between the
spiral angle $\chi $ and the polar angle $\Phi $ in form of the differential
equation

\begin{equation}
\tan \chi =R\deriv{\Phi}{R}\, ,  \label{sechzigdrei}
\end{equation}
where $\tan \chi $ is given by (\ref{sechzig}).

For our further calculations it is convenient to introduce the quantity $%
u=\kappa R.$ Then we arrive at the differential equation

\begin{equation} \label{sechzigvier}
\begin{array}{lll}\bigskip\ds
\abc{a}
\deriv{\Phi}{u} &=& \ds\frac{1}{P}\sqrt{\alpha _{S}+
\frac{A_{Q}\Sigma }{u}+\frac{Q_{1}U(u)}{u^{3}}}\quad\text{with}
 \\ \ds
\abc{b} Q_{1} &=&\ds\frac{3\gamma _{N}M_{c}V(\kappa r_{0})}{r_{0}^{3}}\, ,
\qquad \abc{c}
P=A_{Q}-3u\Sigma \, , \qquad\abc{d}
A_{Q}=\alpha _{Q}\kappa r_{0}\, .
\end{array}
\end{equation}
Integration leads to the implicite form $\Phi =\Phi (R)$ of the
spiral equation $(\Phi _{0}$ constant of integration,
$u_{0}=\kappa r_{0}):$
\begin{equation}
\Phi =\Phi _{0}+
\int\limits_{\xi =u_{0}}^u{
\frac{\sqrt{\alpha _{S}+
\ds\frac{A_{Q}\Sigma }{\xi }+\frac{Q_{1}U(\xi )}{\xi ^{3}}}}{A_{Q}-3\xi
\Sigma }\,\dd \xi}\, .
\label{sechzigfuenf}
\end{equation}
The intervals of the variables are given by
\begin{equation}\label{sechzigsechs}
\begin{array}{lllll}\bigskip
\abc{a} r_{0} &\leqq & R &\leqq& R_{rev}\,,
 \\
\abc{b} u_{0} &\leqq &u &\leqq& \kappa R_{rev} \,.
\end{array}
\end{equation}

For a better astrophysical understanding of the spiral function (\ref
{sechzigfuenf}) and particularly for plotting we are forced to approximate
this integral representation by series expansion about the expansion center
$u_{0}$ according to $\xi =u_{0}+\eta $. Expanding up to the second order in
$\eta $, after a rather lengthy calculation we arrive at the result
($\ds \frac{\eta }{u_{0}}\ll 1)$

\begin{equation}
\Phi -\Phi _{0}=A_{0}\left[ u-u_{0}+\frac{N}{2}\left( u-u_{0}\right) ^{2}
\right]\, ,
\label{sechzigsieben}
\end{equation}
where $A_{0}$ and $N$ are defined by

\begin{equation}  \label{sechzigacht}
\begin{array}{lll} \bigskip
\abc{a}A_{0} &=& \ds
\frac{1}{P_{0}}\sqrt{\alpha _{S}+\frac{A_{Q}\Sigma }{u_{0}}+
\frac{Q_{1}U(u_{0})}{u_{0}^{3}}}\, ,   \\ \bigskip
\abc{b}
N &=& \ds
\frac{3\Sigma}{P_{0}}+\frac{-A_{Q}\Sigma
u_{0}^{2}+Q_{1}u_{0}^{2}\cos u_{0}-3Q_{1}U(u_{0})}{2u_{0}[\alpha
_{S}u_{0}^{3}+A_{Q}\Sigma u_{0}^{2}+Q_{1}U(u_{0})]}\, ,   \\
\abc{c} P_{0} &=& \ds A_{Q}-3u_{0}\Sigma \, .
\end{array}
\end{equation}
Solving the quadratic equation (\ref{sechzigsieben}) we arrive at

\begin{equation}
R=\rho \pm \sqrt{\rho ^{2}-C_{0}}\;\sqrt{1+D(\Phi -\Phi _{0})}
\label{sechzigneun}
\end{equation}
with

\begin{equation}\label{siebzig}
\begin{array}{lll}\bigskip
\abc{a}
C_{0} & = & \ds
\frac{(Nu_{0}-2)u_{0}}{N\kappa ^{2}}\, ,
 \\ \bigskip
\abc{b} D & = & \ds
\frac{2}{A_{0}N\kappa ^{2}(\rho ^{2}-C_{0})}\, ,
\\
\abc{c} \rho & = &
\ds\frac{Nu_{0}-1}{N\kappa }\, .
\end{array}
\end{equation}
This is the explicite form of the spiral equation received by approximation
up to the second order in $\Phi $.

By further power series expansion we get in first order (linearized form)
the Archimedes spiral

\begin{equation} \label{siebzigeins}
\begin{array}{lll}\bigskip
\abc{a} R & = & \ds r_{0}\pm S_{0}(\Phi -\Phi _{0})\quad\text{with}
 \\
\abc{b} S_{0} & = & \ds \frac{1}{2}D\sqrt{\rho ^{2}-C_{0}}\, .
\end{array}
\end{equation}
Hence we find for the distance (at a fixed angle) between two spiral arms
the formula

\begin{equation}
\Delta R=2\pi | S_{0}|\, .  \label{siebzigzwei}
\end{equation}

\section{Numerical evaluation of the theory (model Milky Way)}

Our next aim is to evaluate numerically for an orbiting test body the
general formulas for its radial velocity, azimuthal velocity (rotation
curve) and angular velocity as well as the formula for its spiral motion
(superposition of the radial and azimuthal velocities).

The formulas presented above show that we need two different categories of
parameters, namely first the ``present'' cosmological parameters as
adiabatically fixed parameters determined by our cosmological model which we
treated in our previous papers, and second the individual parameters which
partly are rather well known by empirical observations. Partly we have to
guess two parameters (dark matter parameter $\kappa $, scalaric bremsheat
parameter $\beta _{Q}$) in order to be in rough accordance with the
observations from the motion of the orbiting stars (empirical rotation
curves). Of course, our simplified model can only reflect some basic
features of the astrophysics of galaxies. Specifically our further
application of the above theory refers to a rough model of the Milky Way
(Galaxy).

\subsection{Cosmological and individual parameters}

Numerical cosmological values for the ``present time'':

Here we use our previously published results (Schmutzer 2000b):
\begin{equation}\label{siebzigdrei}
\begin{array}{lll}  \bigskip\ds
\sigma _{c} &=&65.188\, ,   \\ \bigskip\ds \Sigma  & =& 5.54\cdot
10^{-19}\,\mathrm{s}\, , \\ \bigskip\ds \dot{\Sigma} & = &
-3.73\cdot 10^{-36}\,\mathrm{s}^{-2}\, ,
\\ \ds
\alpha_{S} & = & 1.3\cdot 10^{-35}\,\mathrm{s}^{-2}\, .
\end{array}
\end{equation}
Numerical values for rough modelling of the Galaxy:

\begin{equation} \label{siebzigvier}
\begin{array}{llll}\bigskip\ds
M_{c} &=&1.8\cdot 10^{44}\,\mathrm{g}\quad &\text{(acting central
mass)}\, ,
 \\     \bigskip\ds
r_{0} & = & 0.8 \;\mathrm{kpc} &\text{(radius of the inner central
body)}\, ,
\\ \ds
\kappa  &=&0.04\;\mathrm{(kpc)}^{-1}\quad & \text{(dark matter
parameter)}\,.
\end{array}
\end{equation}
Numerical value of the scalaric bremsheat parameter (analog to sun matter,
being explained later):
\begin{equation}
\beta _{Q}=4.39\cdot 10^{3}\, .  \label{siebzigfuenf}
\end{equation}
Hence the parameter (\ref{fuenfzigeins}) and the quantity (\ref
{fuenfzigfuenf}) take the form
\begin{equation} \label{siebzigsechs}
\begin{array}{lll}                                       \bigskip\ds
\alpha _{i} &=&5.89\cdot 10^{-30}\,\mathrm{s}^{-2},   \\ \alpha
_{e} &=&\left(1.302\cdot 10^{-5}+\ds \frac{1.0776\cdot
10^{-3}}{R}\;\mathrm{kpc}\right) \cdot 10^{-30}\,\mathrm{s}^{-2}\,
.
\end{array}
\end{equation}
Concluding this subsection let us add some annotations to the scalaric
bremsheat parameter. According to its definition (\ref{vierzigsechs}) it
mainly depends on following values of the inner central body of the Galaxy:
moment of inertia $J$ (\ref{zwanzigsieben}), cubic heat expansion
coefficient $\alpha _{c}$ and heat capacity $W$ (\ref{zwanzigdrei}) which
contains the specific heat $c_{Q}$ of the central body. The heat consumption
factor $f_S$ seems to be of the order of magnitude 1. Let us
further mention that we know the radius $r_{0}$ of the central body and
roughly its angular velocity $\omega $. Further we realize that the mass of
the central body cancels. Therefore we are mainly left with the parameters
$\alpha _{c}$ and $c_{Q}$ . Let us for numerical estimates tentatively use
the values for the sun, being available in astrophysical literature:
\begin{equation}
\alpha _{c\mathrm{sun}}\approx \mathrm{3.66\cdot 10^{-3}K}^{-1}\, ,
\quad
c_{Q\mathrm{sun}}\approx  1.24\cdot 10^{8}\,
\mathrm{cm^{2}\, s^{-2}\, K}^{-1}\,.
\label{siebzigsieben}
\end{equation}
Hence we obtain for the sun the value
$\beta _{Q_\mathrm{sun}} \approx 4.39\cdot 10^{3}$.
This is our explanation for our previous choice of the
parameter $\beta _{Q}$.

\subsection{Plotting}

The above considerations led us to the parameters needed for plotting, i.e.
now we are able to plot the radial velocity curve (\ref{vierzigfuenf}) and (%
\ref{fuenfzigacht}), the rotation curve (\ref{zehnsieben}) and (\ref
{zehnneun}), and the angular velocity curve (\ref{zehnacht}) and (\ref
{zwanzig}). Fig.1 and Fig.2 show the radial course of the radial velocity $%
V_{R}$ and the azimuthal velocity $V_{\Phi }$ (abscissa $R$ in kpc,
ordinates $V_{R}$ and $V_{\Phi }$ in km s$^{-1}).$ Fig. 3 presents the
radial course of the angular velocity $\omega $ in s$^{-1}.$

\begin{figure}[p]
\begin{center}
\footnotesize
\includegraphics[width=0.75\columnwidth]{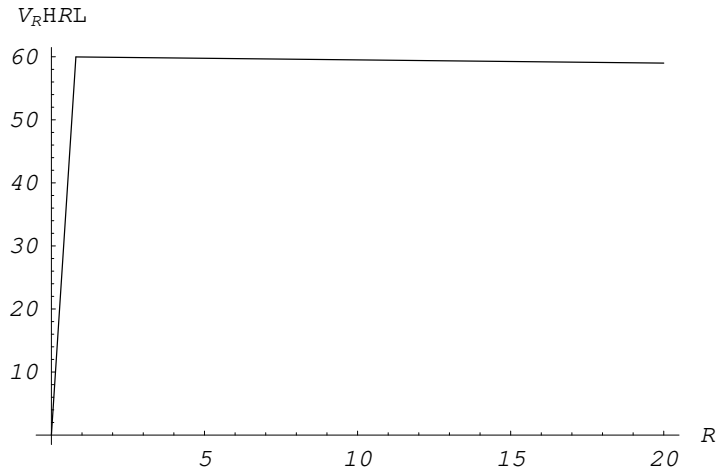}  
\caption{Radial course of the radial motion of the star}
\label{fig:1}
\end{center}
\end{figure}

\begin{figure}[p]
\begin{center}
\footnotesize
\includegraphics[width=0.75\columnwidth]{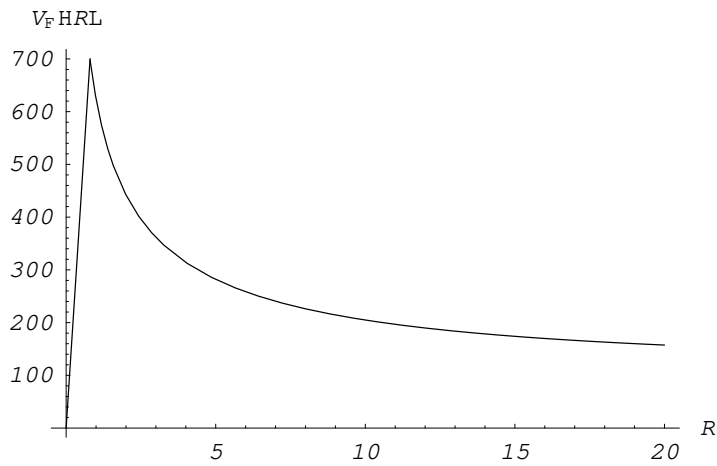}  
\caption{Radial course of the
azimuthal velocity of the star (rotation curve)}
\label{fig:2}
\end{center}
\end{figure}

\begin{figure}[p]
\begin{center}
\footnotesize
\includegraphics[width=0.75\columnwidth]{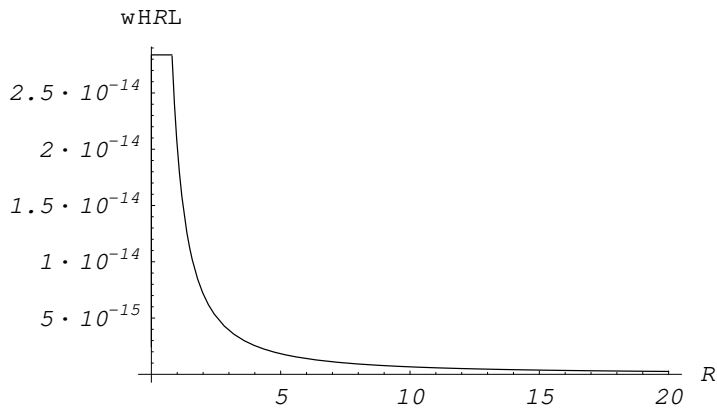}  
\caption{Radial course of the
angular velocity}
\label{fig:3}
\end{center}
\end{figure}

Plotting of the spiral of the orbiting body is based on the formula (\ref
{sechzigneun}) containing the parameters $\rho$, $C_{0}$ and $D$, which after
some substitutions are determined by the equations (\ref{siebzig}). A rather
lengthy numerical calculation leads to the values

\begin{equation} \label{siebzigacht}
\begin{array}{lll}\bigskip
\rho &=& 1.33\,\mathrm{kpc\, , }   \\ \bigskip
C_{0} &=&1.49\,\mathrm{ (kpc)}^{2},  \\
D &=&-0.259\,.  \nonumber
\end{array}
\end{equation}

Fig.4 shows the spiral, where the radial distance is given in kpc.

\begin{figure}[p]
\begin{center}
\footnotesize
\includegraphics[width=0.65\columnwidth]{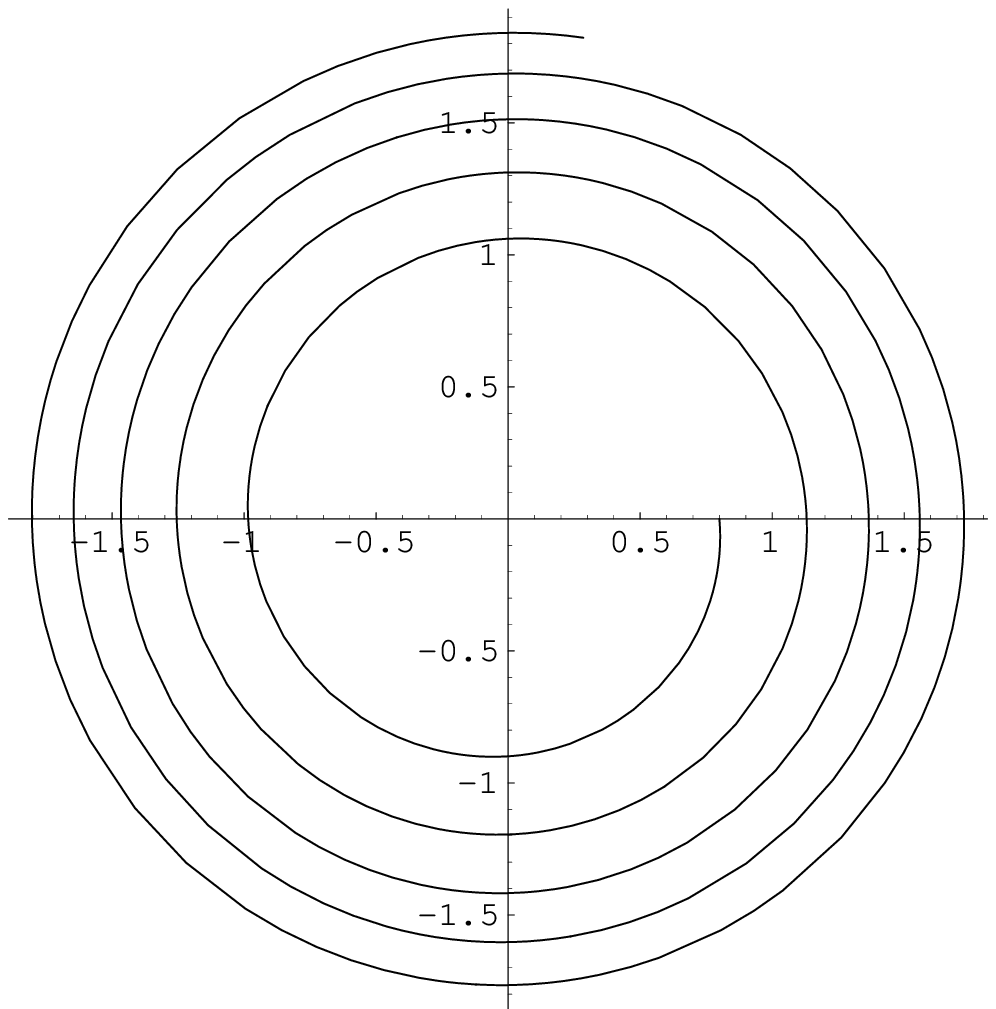}  
\caption{Spiral motion of the orbiting star}
\label{fig:4}
\end{center}
\end{figure}

 \vspace{2ex}

I would like to express my thanks to Prof. Dr. A. Gorbatsievich
(University of Minsk) for scientific discussions and to Mr. E.
Schmidt (Jena) for technical advice.\\[2ex]

\noindent \textbf{References}
\begin{description}
\item[]
Schmutzer, E.: 1995, Fortschritte der Physik 43, 613
\item[]
Schmutzer, E.: 2000a, Astr. Nachr. 321, 137
\item[]
Schmutzer, E.: 2000b, Astr. Nachr. 321, 209
\item[]
Schmutzer, E.: 2000c, Astr. Nachr. 321,227
\item[]
Schmutzer, E.: 2001, Astr. Nachr. 322 (in press)
\item[]
Dehnen, H. , Rose, B. and Amer, K. : 1995, Astrophysics and Space
Science 234, 69
\end{description}

\vspace{4ex}

\noindent Address of the author:\\[2ex] Ernst Schmutzer Cospedaeer
Grund 57 D-07743 Jena Germany

\end{document}